%% file: main_arxiv.tex
    \documentclass{nst}
    \usepackage{csquotes}
    \usepackage{xcolor}
    \usepackage{multirow}
    \usepackage[english]{babel} 
    
    \usepackage{siunitx}
    \DeclareSIUnit{\gram}{g}
    \DeclareSIUnit{\mol}{mol}
    \DeclareSIUnit{\electronvolt}{eV}
    
    \definecolor{color1}{RGB}{0,0,90} 
    \definecolor{color2}{RGB}{0,20,20} 
    
    \usepackage{graphicx}
    \usepackage{hyperref} 
    \usepackage{wasysym} 

    \hypersetup{
    	hidelinks,
    	colorlinks,
    	breaklinks=true,
    	urlcolor=color2,
    	citecolor=color1,
    	linkcolor=color1,
    	bookmarksopen=false,
    	pdftitle={Title},
    	pdfauthor={Author},
    }

\begin{document}

    \title{Full Characterization of a Mock Nuclear Waste Barrel with Muon Tomography using Micromegas Detectors}

    \thanks{Review version}%

    \author{Raphaël Bajou}
    \email{r.bajou2$@$gmail.com}
    \affiliation{CEA/Irfu, Université Paris-Saclay, 91191 Gif-Sur-Yvette, France}
    \author{David Attié}%
    \affiliation{CEA/Irfu, Université Paris-Saclay, 91191 Gif-Sur-Yvette, France}%
    \author{Héctor Gómez}
    \affiliation{CEA/Irfu, Université Paris-Saclay, 91191 Gif-Sur-Yvette, France}%
    \affiliation{Centro de Astropartículas y Física de Altas Energías (CAPA), Universidad de Zaragoza, 50009 Zaragoza, Spain}
    \author{Irakli Mandjavidze}
    \affiliation{CEA/Irfu, Université Paris-Saclay, 91191 Gif-Sur-Yvette, France}%
    \author{Philippe Mas}
    \affiliation{CEA/Irfu, Université Paris-Saclay, 91191 Gif-Sur-Yvette, France}%

    \date{\today}

    \begin{abstract}
    Muon tomography based on multiple Coulomb scattering provides a non-destructive method to image dense and shielded objects using naturally occurring cosmic-ray muons. In the context of nuclear waste characterization, we present the experimental imaging of a 205-L mock waste barrel using a dedicated \SI{1}{\meter\squared} muon scattering tomography test bench. The system employs multiplexed resistive Micromegas detectors, enabling stable and high-precision muon tracking. 
    Monte Carlo simulations are first used to characterize material-dependent scattering signatures and to quantitatively assess identification performance using statistical reconstruction. These simulation-based results are then used to define objective discrimination thresholds, which are subsequently applied to experimental data for the localization and identification of internal anomalies. Using an Angle Statistics Reconstruction algorithm, we achieve a spatial resolution of 10 mm and demonstrate the three-dimensional imaging of  an internal structure containing both low- and high-radiation length materials.
    Material discrimination performance is evaluated using receiver operating characteristic analysis, yielding high identification efficiency for dense metallic inclusions such as lead and steel (AUC $\geq$ 0.96) within acquisition times of a few days, while cavities also exhibit strong contrast. Experimental results show good agreement with detailed Monte Carlo simulations. By establishing a continuous workflow from simulation-based performance characterization to practical application on measured data, this work provides a quantitatively validated framework for muon scattering tomography applied to complex, shielded objects.
    
    \end{abstract}


    \maketitle 
    
    \tableofcontents 
    
    \thispagestyle{empty} 
    
    
    \section*{Introduction} 
    
    \addcontentsline{toc}{section}{Introduction} 
    
    Muon tomography has emerged as a non-invasive imaging technique that exploits the natural flux of cosmic-ray muons to probe the internal structure of objects that are otherwise difficult or impossible to inspect. Owing to their high penetration capability, muons enable the investigation of dense and shielded volumes without the need for artificial radiation sources. Over the past decade, muon-based imaging has been applied in a variety of contexts, including nuclear inspection \cite{Lefevre2025}, archaeology \cite{Morishima2017, Procureur2023b}, geoscience \cite{Lechmann2021, Bajou2023, Tramontini2024, Balazs2025}, and industrial monitoring \cite{Arbol2022}, illustrating its broad applicability across different environments and length scales.

    Among the different muon imaging configurations, multiple Coulomb scattering tomography relies on the angular deflections experienced by muons as they traverse matter to infer material properties and internal structures. The performance of scattering tomography depends on several factors, including the spatial resolution and stability of the muon-tracking detectors, as well as the reconstruction algorithms used to extract scattering information. In this context, Micromegas detectors—particularly resistive multiplexed designs—offer high spatial resolution, good rate capability, and reduced sensitivity to X-ray and gamma backgrounds, making them well suited for muon tracking in environments relevant to nuclear applications.

    Feasibility studies and three-dimensional reconstruction approaches in muon scattering tomography (MST) are primarily developed and evaluated using Monte Carlo simulations, including statistical reconstruction methods such as Maximum Likelihood Expectation Maximization (MLEM) \cite{Schultz2007}, Angle Statistics Reconstruction technique (ASR) \cite{Stapleton2014}, or more recently in \cite{ughade_trec_2025}. These simulation-based approaches are essential for developing reconstruction algorithms and for quantifying material discrimination performance under controlled conditions. However, the translation of such performance metrics into practical decision criteria applicable to experimental data—such as the definition of identification thresholds and their use for anomaly localization—has received comparatively less attention. Experimental studies that establish a clear methodological continuity between simulation-based performance characterization and its application to real measurements on complex, heterogeneous objects therefore remain particularly valuable.

    In this study, we present an experimental evaluation of a realistic mock nuclear waste barrel using MST. A dedicated 1 m$^2$ cosmic-muon detection bench equipped with resistive multiplexed Micromegas detectors is used to image a 205 L metallic barrel filled with concrete and containing a variety of low- and high-radiation-length materials. Using an ASR algorithm \cite{Stapleton2014}, Monte Carlo simulations are first employed to quantify material identification performance and to define objective discrimination thresholds. These thresholds are then applied to experimental data to guide the localization and identification of internal anomalies. The spatial resolution and discrimination capability of the system are assessed as a function of exposure time, and experimental results are systematically compared with simulation predictions. This combined simulation-to-data approach aims to provide a practical and quantitatively grounded methodology for MST applied to complex, shielded objects.
    
    
    
    \section{Multiple scattering tomography}
    
    \subsection{Multiple Coulomb scattering}
    As charged leptons, muons interact with the nuclei electromagnetic field, resulting notably in a sum of small angular deflections along their path in matter. This succession of Coulomb scatters is parametrized by a scattering angle $\theta$ whose distribution features long tails due to less-frequent hard scatters but is overall 98\% Gaussian \cite{ParticleDataGroup2024}:   
    \begin{equation}
       P(\theta)d\Omega   = \frac {1} { 2\pi \theta_0 ^2 }   \exp \left ( - \frac{\theta ^2} {2\theta_0 ^2} \right ) d\Omega
    \end{equation}
    where $\Omega$ is the solid angle and $\theta_0$ the width of the distribution. The latter depends on the particle kinematics and propagation medium properties: 
    \begin{equation}
    \label{eqTheta0}
         \theta_0 \simeq   \frac{13.6} { \beta c p } \sqrt\frac{X} {X_0} \left ( 1 + 0.038 \ln \left ( \frac {X}{X_0 \beta ^ 2}  \right)\right )
    \end{equation}
    where  $\beta c$ is the particle velocity, $p$ the momentum in \si{\giga\electronvolt}, $X$ the thickness of the crossed material, and $X_0$ its radiation length estimated from the density $\rho$, atomic mass $A$, number $Z$ with the following expression:   
    \begin{equation}
        X_0 = 716.4 \text{g.cm}^{-2} \frac {A} {\rho Z(Z+1) \ln \left ( \frac {287}  {\sqrt {Z}} \right ) }
    \end{equation}
    The overall deflection magnitude represented by the angular width $\theta_0$ in Eq. \ref{eqTheta0} is at first order both inversely proportional to momentum $p$ and $\sqrt{X_0}$. Consequently, the abundance of low-energy muons (below 10 GeV) in the flux received at Earth's surface, combined with the minimal energy loss mainly through ionization at this energy range, makes the multiple Coulomb scattering (MCS) of peculiar interest for detecting high-$Z$ crossed material, typically with short radiation length $X_0$.  
    
    The first proofs-of-concept regarding the use of muons MCS for tomography were released at LANL in the early 2000s \cite{Borozdin2003, Schultz2004}, targeting special nuclear material (SNM) identification inside vehicles and shipping containers in the context of border security control. 
    
    More recently, developments were initiated to assess the potential of MST for nuclear waste characterization with the scope of deploying a scanning system in storage facilities to perform non-destructive waste containers inspection. The muon imaging of a 500-liter stainless-steel intermediate level waste (ILW) drum was published in 2018 by Mahon et al. \cite{Mahon2018}, containing uranium and lead samples. The imaging reached up to 10 mm resolution in the shown horizontal XY slices obtained for a total record of 2.47$\times10^{7}$ muon events.  
    This study set a benchmark for the work presented hereafter, which has led to the imaging of a smaller 205-L mock waste barrel containing a larger range of low and high-$Z$ objects for a comparable muon exposure. We described in the following sections the characteristics of the detection system used for this.

    \subsection{MultiGen detector}
    \label{sec:MultiGen}
    To detect the passage of cosmic muons, the gaseous detection technology Micromegas (\cite{Giomataris1996}, \cite{He2023}) was employed in a MultiGen detector design. 
    A MultiGen is composed of a resistive multiplexed Micromegas with a 2D layout with an active area of 50$\times$50~cm$^2$ \cite{Bouteille2016a}. The multiplexing feature allows for efficient readout of multiple channels using fewer electronics, reducing complexity and cost. This design was tested first in the WatTo experiment \cite{Bouteille2016b} using four prototypes of the MultiGen detector, where the first outdoor telescope was operated for 3.5 months, including a 1.5-month period powered by solar panels, demonstrating its low-power consumption of approximately 30~W. 
    The last version of MultiGen used in this study features 732 strips in each coordinate (X \& Y) read out by 61 out of 64 channels of the Dream chip of the Front Unit Card.
    
    \subsection{Detection setup}
    \label{sec:Setup}
\begin{figure}[!ht]
    \includegraphics[width=\linewidth]{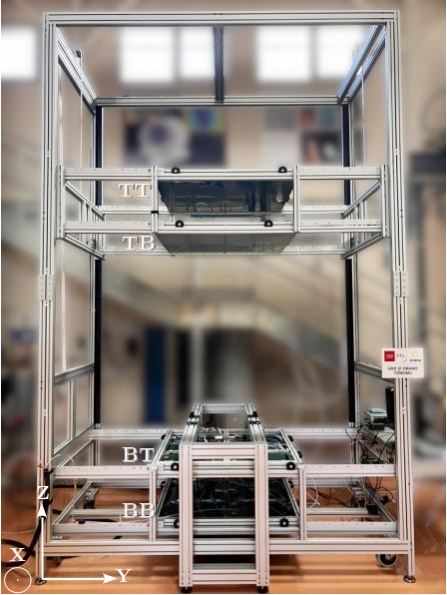}
    \caption{Cosmic muon detection bench at CEA/Irfu, featuring the top and bottom trackers, respectively composed of the TT-TB and BT-BB detection panel pairs. Each panel is composed of four MultiGen detectors. The barrel support table is overhanging the bottom tracker.}
    \label{fig:bench}
    \end{figure}

     The cosmic muon bench is composed of a top and a bottom tracker, shown in Fig. \ref{fig:bench}. Both trackers feature two horizontal detection panels, vertically spaced by \SI{40}{\centi\meter}. From top to bottom, the panels are respectively labelled TT, TB, BT, and BB. Each panel gathers four MultiGen detectors (cf. Sec. \ref{sec:MultiGen}), providing a total 1 m$^2$ active area. 
     The mechanical structure hosts a support table made of aluminium frames to position the object to scan, in addition to a pulley system to move vertically along the Z-axis the top tracker and thus adapt the spacing between top and bottom modules to the height of the scanned object. In the present configuration, the two trackers are spaced by \SI{1.36}{\meter}. 
    
    \subsection{Mock barrel}
    \label{sec:Barrel}
    For this study, a \SI{205}{\liter} metallic barrel with inner diameter \SI{570}{\milli\meter} and height \SI{800}{\milli\meter} was selected. The barrel is a non-alloy steel cylindrical drum with a wall thickness of 2 mm, closed by two end caps of the same thickness. This type of container is for instance used to store cemented sludge and concentrates waste (type F3-6-02 according to National Agency for Radioactive Waste Management classification).
    
    The empty barrel was filled with poured concrete of density $\rho =$ \SI{2.1}{\gram.\centi\meter^{-3}} containing a collection of nine objects including six types of material: graphite (neutron-moderating material), polyvinyl chloride (PVC), water, air, lead, and steel. The PVC material corresponds to an empty plastic bottle manually filled with PVC laboratory gloves, for a density estimated between 0.7 and 0.9  \si{\gram.\centi\meter^{-3}}, hence lower than the reference density $\rho_{\text{NIST}} = 1.3$ \si{\gram.\centi\meter^{-3}}. 
    
\begin{figure}[ht]\centering 
    	\includegraphics[width=\linewidth]{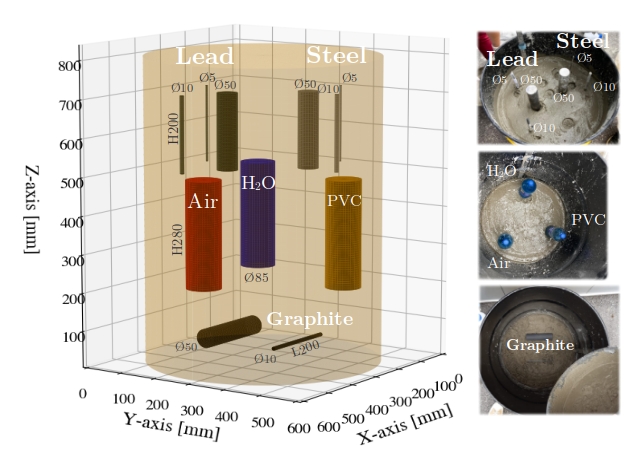}
    	\caption{Scheme of the objects distribution inside poured concrete with corresponding dimensions alongside the photos taken at the three height levels during the barrel assembly.}
    	\label{fig:barrelcontent}
    \end{figure}

    The objects were placed at three different heights as shown in Fig. \ref{fig:barrelcontent}, approximately ranging along the Z-axis in [60, 110], [220, 500], and [550, 750] mm. 
    
    \section{Data acquisition and reconstruction}
    \label{sec:Reconstruction}
    The barrel described in Sec.\ref{sec:Barrel} was installed in vertical position from early September 2024 to the end of November 2024. During this period, 19 data runs were taken for a total acquisition time of nearly 52 days, recording up to $2.34 \times 10^7$ events, at a mean trigger rate of 5.2 Hz.
    
    During acquisition, an event is recorded when a particle is synchronously detected in the top and bottom trackers, crossing all four detection panels  in a \SI{2}{\micro\second} time acquisition window. 
    One potential source of contamination is the accidental combinatorial coincidence of two independent muons crossing the top and bottom trackers within the same time window, which would be reconstructed as a single, spurious event. Assuming perfect detection efficiency and independent muon arrivals following a $\cos^2{\theta}$ angular intensity (with vertical intensity $I_0 \simeq$  \SI{70}{\meter^{-2}\second^{-1}\steradian^{-1}} \cite{DePascale1993}), the geometrical acceptance of the four-panel bench yields an estimated open-sky four-panel bench rate of about \SI{12.4}{\hertz}, against a two-panel tracker rate at \SI{71.7}{\hertz}, therefore \SI{59.3}{\hertz} outside the four-panel acceptance. The resulting accidental coincidence rate is then estimated as \SI{1.41e-2}{\hertz}, corresponding to about 0.11\% of the theoretical four-panel event rate. As this estimate assumes perfect and spatially homogeneous detection efficiency across all panels, it is regarded an upper bound on the combinatorial contamination. This combinatorial background is not reproduced in the GEANT4 simulation (Sec.\ref{secSimulation}), which generates single, independent muon tracks.
    The two trackers each provide a track segment from the chi-squared fit of strip clusters formed on MultiGen detectors (cf. Sec.\ref{sec:MultiGen}). 
    These top and bottom segments are respectively extrapolated to incident and outgoing rays inside the volume-of-interest (VOI), divided into cubic voxels. In a Cartesian coordinate system, the two ray directions give a first view of the scattering geometry, giving the total scattering angle $\theta$ in space.
     Furthermore, particle scattering can be studied in the two transverse directions, orthogonal to the propagation direction  \cite{ParticleDataGroup2024}. A common practice to collect twice as much information about the event is to consider the projected angles taken as independent random variables identically distributed in the XZ and YZ planes, with the Z-axis as the propagation axis. In the small angles approximation, the deflection angle $\theta$ in space can be decomposed as $\theta^2 \simeq \theta_x ^2 + \theta_y ^2$, where $\theta_x$ and $\theta_y$ are the projected planar angles in XZ and YZ.
    The goal of the reconstruction is then to associate these estimates, derived into a scattering score, to the appropriate voxels.
    
    A primary reconstruction, first introduced for MST in \cite{Schultz2004}, consists of assimilating all the small deflections undergone by the muon in matter into a single scattering vertex. This unique position and its corresponding voxel can thus be geometrically determined by taking the minimal orthogonal distance between the two rays, where lies the point-of-closest-approach (PoCA) at the centre of the segment connecting the pair of closest ray points.
    This method represents an analysis starting point offering a fast event reconstruction but suffers from the over-simplified unique vertex assumption, which neglects the stochastic nature of MCS, and is put into defect when rays do not intersect in the VOI, resulting in a significant loss fraction of events (cf. Sec. \ref{secDiscussion}).   
    
\begin{figure}[ht]\centering 
    	\includegraphics[width=\linewidth]{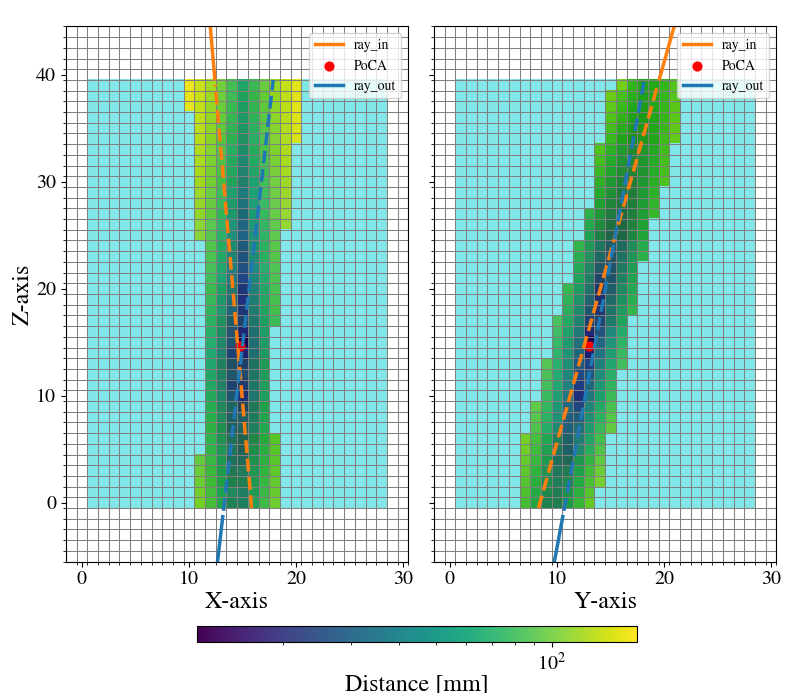}
    	\caption{Two-dimensional displays of a scattering muon event in barrel VOI (light blue voxels) projected in XZ and YZ planes. The displays feature the incident (in orange) and outgoing (in dark blue) rays,  their PoCA, and the voxel distance to rays as the third coordinate in a coloured map. The axes are in voxel units (here 1 voxel = 20 mm).}
    	\label{fig:evd2d}
    \end{figure}

    To go beyond the single voxel scoring of the PoCA method, a reconstruction method adapted from the \textit{Angle Statistics Reconstruction}, introduced in \cite{Stapleton2014}, was implemented to better distribute the scattering score estimate along the voxels traversed by muon rays in the VOI.  
    This process consists of distributing for each muon $i$ not one but two score values, i.e. both projected scattering angle absolute values $|\theta_{i,x}|$ and $|\theta_{i,y}|$, along the voxels that satisfy a distance criterion to both incident and outgoing rays. As no momentum estimate is available with the present configuration, the score is directly computed from the projected angles, assuming a constant reference momentum $p_0=1$\si{\giga\eV\per c} similar to the original ASR formulation in \cite{Stapleton2014}.
    To identify which voxels to score, the following quantity is introduced:
    \begin{equation}
        D_{ij} = \max(\min_{z}(||r_{\text{in}, i}(z) - c_j||), \min_{z}(||r_{\text{out}, i}(z) - c_j ||)) 
        \label{eq:MaxDistanceMuonVoxel}
    \end{equation}
    where $r_{\text{in}, i}(z)$ and $r_{\text{out}, i}(z)$ are respectively the incident and outgoing rays of the $i$-th muon, and $c_j$ the voxel centre coordinate of the $j$-th voxel. 
    $D_{ij}$ corresponds to the larger of the two minimal distances to the incident and outgoing rays, so that a voxel $j$ is scored only if it lies within a distance $d_{th}$ of both segments of the muon estimated trajectory:
    \begin{equation}
    D_{ij} \leq d_{th}
    \label{eq:DistanceCriterion}
    \end{equation}
    The distance threshold $d_{th}$ is typically set to the voxel size. To illustrate the process, Fig.\ref{fig:evd2d} shows an example with voxel size set to 20 mm of a simulated event projected in both deflection planes, displaying the distance $D_{ij}$ to incident and outgoing rays for neighbouring voxels. On both two-dimensional displays, the coloured pixels are the ones in a volume with a proximity of three voxels to either incident or outgoing rays. Here, the pixel values are the minima recorded along the third orthogonal axis (directed out of the page).
    
    After processing all events, the final score model is obtained voxel-wise by taking the median or the relevant quantile of the pooled distribution of $|\theta_{x}|$ and $|\theta_{y}|$ obtained from processed muons satisfying the distance criterion (Eq.\ref{eq:MaxDistanceMuonVoxel} and \ref{eq:DistanceCriterion}). As noted in \cite{Stapleton2014} and observed during data reconstruction, the 75th percentile gives the optimal contrast in the model for material identification within the concrete barrel presented in \ref{secResults}.
    
    In the next section, we present and compare the results obtained with this reconstruction method from real and simulated data.
    
    
    \section{Results and Discussion}
    
    \subsection{Results on real data}
    \label{sec:ResultsRealData}
\begin{figure}[ht]
        \centering
        \includegraphics[width=\linewidth]{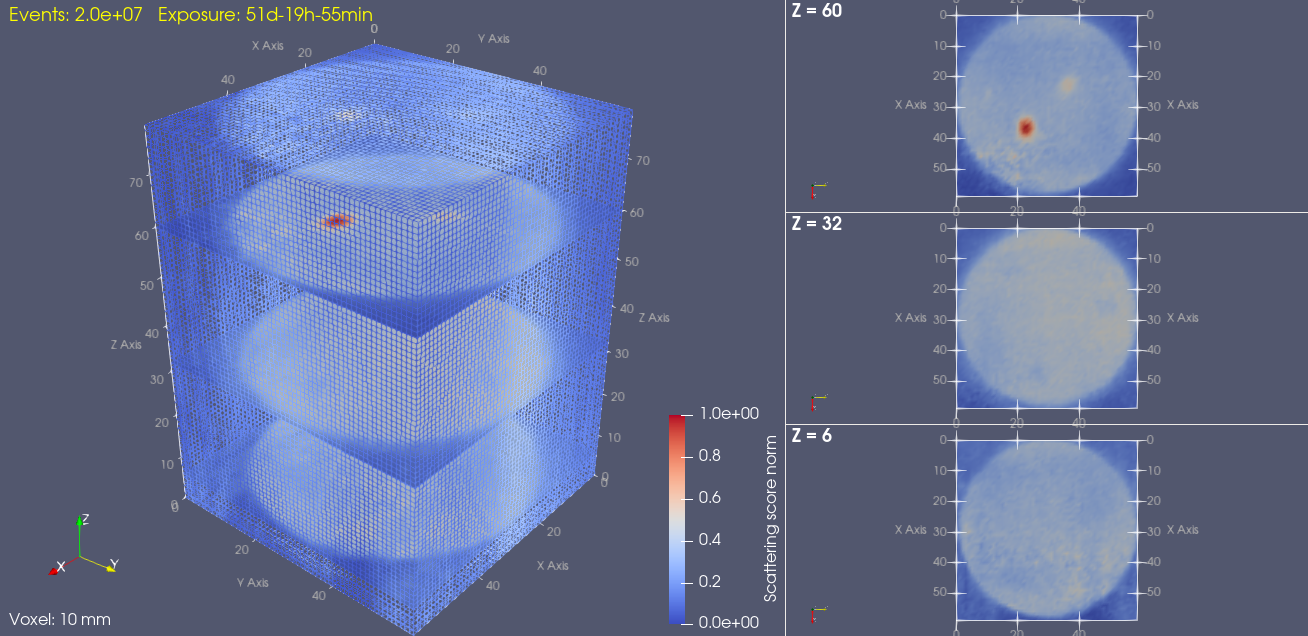}
        \caption{(left) Three-dimensional wireframe view of the scattering score model (min-max scaled) obtained from $2.0\times10^7$ real muon events with voxel size set to \SI{10}{\milli\meter}; 
        (right) Three horizontal XY slices taken inside the 3D model at respectively $\text{z}\in[6,32,60]$ in voxel unit.}
        \label{fig:asr_3dmodel_real}
    \end{figure}

\begin{figure}[ht]
        \centering
        \includegraphics[width=\linewidth]{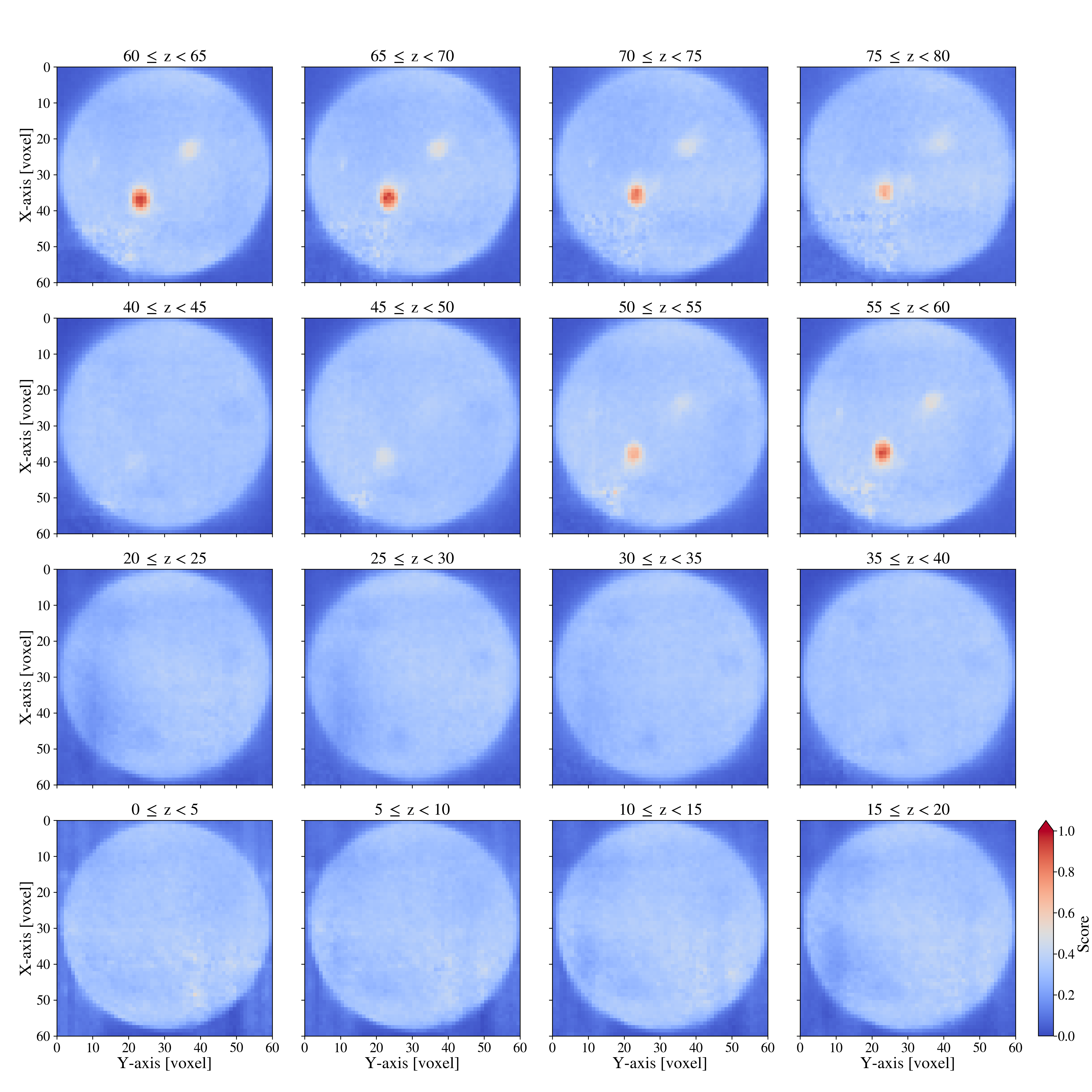}
        \caption{Scattering score images obtained for real data with $2.3\times10^7$ events and voxel size set to \SI{10}{\milli\meter}. Each image contains the mean score obtained on the given 50mm-thick horizontal slice in the VOI.}
        \label{fig:score_real}
    \end{figure}

    For this study, the VOI $60\times60\times80$ \si{\cubic\centi\meter} was divided into $2.88\times10^5$ voxels of 1x1x1\si{\cubic\centi\meter}.  The selection of scattered events with $\theta \geq 1$ \si{\milli\radian} ($\sim$ angular resolution of the MultiGens) leads to a three-dimensional scattering score distribution. Accordingly to the cylindrical geometry and vertical positioning of the studied object, this 3D distribution can then be sliced horizontally for a better rendering of the contrasts created by the different materials, as displayed in Fig.\ref{fig:asr_3dmodel_real}. 
    In Fig.\ref{fig:score_real}, the 3D distribution is divided into sixteen two-dimensional images. Each image is a horizontal slice in XY plane with a 50-mm thickness range along the Z-axis, encompassing a layer of five voxels. The coordinates are given in voxel units in the VOI referential. Each pixel then contains the mean score estimated from the scores of those five-layered voxels. 
    As the thickness of the crossed concrete material decreases with height z, the concrete score value also decreases with z. To cope with this effect, the distribution was normalized along the Z-axis by the score's most probable value estimated in each XY projection. 
    
    \label{secResults}
    
    In height $\text{z}\in[50; 75]$, the \O50-\si{\milli\meter} lead and steel vertical bars visually stand out at first glance from the surrounding concrete in the barrel as contrasting spots, located respectively at the bottom left for the reddest spot and top right of the image centres. 
    With a less pronounced contrast, the \O5-\si{\milli\meter} lead bar is also spotted between $\text{z}\in[55; 70]$ at the centre left of the images around (x,y) = (26,10) voxels. 
    In the top half of the barrel, i.e., between $\text{z}\in[40; 80]$, a blurring effect is to be seen at the bottom left corner of the images. This effect is expected to be caused by the performances of the two MultiGen detectors in the bottom-left quarter of the top tracker, whose detection efficiency is globally reduced on their whole detection surface by a factor of two compared to other detectors on the same panels. 
    The bottles forming large $X_0$ sub-volumes (water, PVC, air) are best visible from $\text{z}\in[20; 45]$. On the five concerned images, the bottle of water is retrieved at the top left, the PVC one at the bottom centre, and the air at the centre right. These images are notably affected at $\text{z}\in[20; 35]$ by a large vertical artefact spanning from the centre left to bottom left and causing an additional cavity score anomaly. This is attributed to the presence of a lower efficiency region on the bottom-left detector on the bottom-top panel. On MultiGens, these detection inefficiencies are due to clusters of dead or noisy strips in the X or Y direction. 
    Lastly, at the lowest heights between $\text{z}\in[0; 20]$, the inefficiencies reflect again as score inhomogeneities and vertical band-shaped artefacts, whereas the scattering score would be expected to be uniformly distributed in the barrel volume due to both minimal density and atomic number contrasts between the two graphite bars and the surrounding concrete. 
    These two artefacts (blurring in $\text{z}\in[40; 80]$ and vertical band-shaped), although both rooted in detector-level inefficiencies, arise from distinct mechanisms: the blurring effect results from a global, factor-of-two reduction in detection efficiency over the full surface of the affected top-tracker modules, which increases statistical noise in the reconstructed score without introducing a specific geometric pattern. In contrast, the vertical band-shaped artefact originates from localized clusters of dead or noisy readout strips on a single bottom-tracker detector, which locally suppress or corrupt the muon hit information along well-defined X or Y coordinates, producing a spatially structured artefact.
    Besides these effects, in the outer barrel region, slight contrasts appear along the X-axis on the vertical edges of the images created by the two aluminium frames of the support table overhanging the bottom tracker (cf. Fig.\ref{fig:bench}).

    \subsection{Comparison with simulation}
    \label{secSimulation}
    The experimental results were compared to a simulation setup in GEANT4 \cite{Agostinelli2003}, a particle propagation Monte-Carlo simulation toolkit, reproducing the MCS inside the barrel and the expected muon signal in cosmic bench detectors, each with perfect efficiency. The muons are generated according to the Guan flux parameterisation \cite{Guan2015}. 
    
\begin{figure}[ht]
        \centering
        \includegraphics[width=\linewidth]{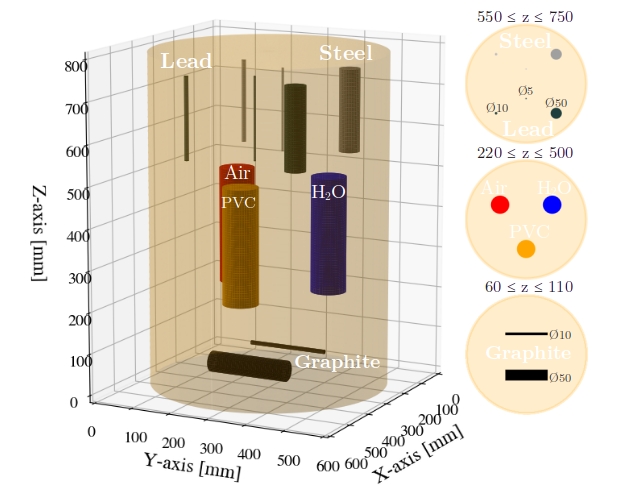}
        \caption{Scheme of the objects distribution in the GEANT4 simulation. The dimensions of the cylindrical objects are identical to the real objects in the mock barrel Fig.\ref{fig:barrelcontent}.}
        \label{fig:barrel_content_simu}
    \end{figure}

    In Fig.\ref{fig:barrel_content_simu}, similarly to the real mock barrel described in Sec.\ref{sec:Barrel}, the simulated objects are staged upon three height levels. The barrel metallic shell is a \SI{2}{\milli\meter} thick iron cylinder. All the inner objects are reproduced with the same materials and dimensions (radius and height), only their XY centre positions differ from the real objects.

\begin{figure}[ht]
        \centering
        \includegraphics[width=\linewidth]{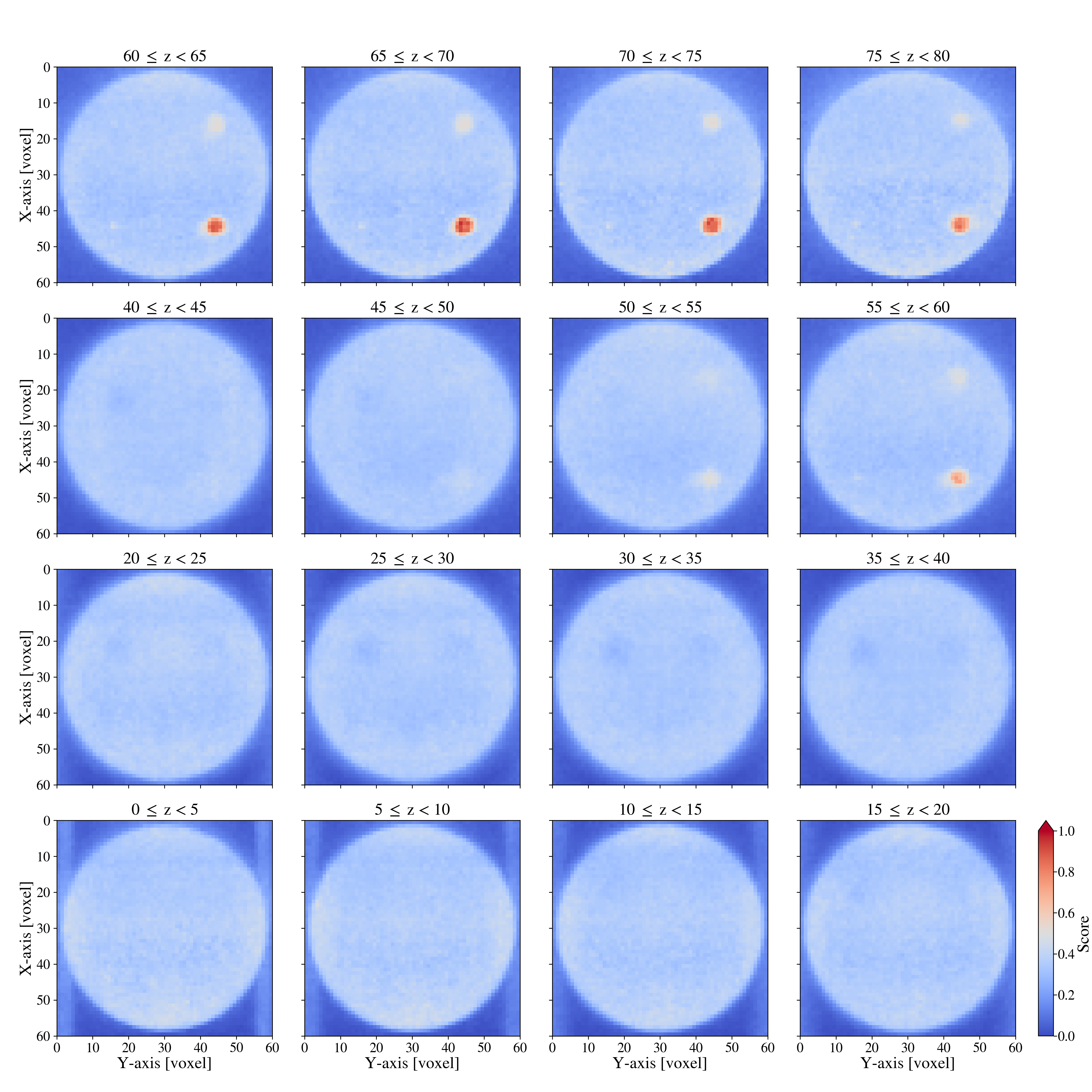}
        \caption{Scattering score images for simulation data obtained from $7.7\times10^6$ muons with voxel size set to \SI{10}{\milli\meter}. }
        \label{fig:score_simu}
    \end{figure}

    Similarly to real data, the scattering score modelling on simulation data is presented in Fig. \ref{fig:score_simu}. The distribution was normalized as described in Sec.\ref{sec:ResultsRealData} to compensate for the score dependency with concrete vertical thickness. As a result, the object contrasts are very similar to the real data model. The aluminium frames of the barrel support table, visible at $\text{z}\in[0; 15]$, and the mechanical structure holding detectors together on the tracker panels were reproduced in the simulation. Though some differences are observed: 
    the \O5-mm lead bar is not visible in the z $\in[55;75]$ range, while the \O10-mm lead bar is here identified due to the absence of detection blurring evoked in Sec.\ref{sec:ResultsRealData}; the simulated PVC bottle in the bottom centre in images z $\in[20;50]$ is less distinguishable than in the real model, likely due to its material of nominal PVC density, higher than in the real assembled object as mentioned in Sec.\ref{sec:Barrel}, which at the end lowers the contrast to concrete. Furthermore, in the bottom four images, a slight cross-pattern artefact caused by the separation of intra-detectors (46 mm wide both in the X and Y-axis) is highlighted by the absence of detection effects. This subtle artefact, visible only in simulation, is masked in real data by the larger-amplitude artefacts arising from detector inefficiencies (cf. Sec. \ref{sec:ResultsRealData}), which dominate the local score variations at a comparable spatial scale.
    
    \subsection{Material identification performance}
    \label{sec:RocStudy}
    
    The goal of this work was to assess the ability of the current MST bench and reconstruction method described in Sec.\ref{sec:Reconstruction} to detect both low-$X_0$ and high-$X_0$ objects within the surrounding concrete in the scattering score model. In this model, the different materials schematically represent anomalies of two types: inclusions (lead and steel bars) and cavities (air, water, PVC bottles), causing respectively local score maxima and minima as seen in Fig.\ref{fig:score_real} and Fig.\ref{fig:score_simu}. The graphite bars constitute the most challenging type of material detection here given its $X_0$ of \SI{19.4}{\centi\meter} among the concrete with $X_0\sim$\SI{12.6}{\centi\meter} at \SI{2.1}{\gram.\centi\meter^{-3}} \cite{ParticleDataGroup2024}, in addition to their horizontal placement at the bottom of the barrel, shielded by more than 70 cm of concrete. 
    
    For this purpose, the Receiver Operating Characteristics (ROC) study provides a standard framework \cite{Fawcett2006} to quantify the discrimination between a given object and concrete voxels, treated as a binary classification problem where score estimates are compared voxel-wise to a certain threshold value and give either a positive or negative detection. Hence, the ROC relies on the estimation for all threshold values of the True Positive Rate (TPR) or "sensitivity" and the True Negative Rate (TNR) referred to as "specificity". The ROC curve renders all the possible realisations in the sensitivity versus specificity complementary, i.e., the False Positive Rate (FPR), space. The area under the curve (AUC) obtained by integration provides a standard classification performance metric. The closer the AUC gets to 1, the better the separation between the positive and negative voxel groups. To the contrary, an AUC of 0.5 shows that the voxel classification in that case is equivalent to random guessing. This lower limit is represented by a diagonal line from (0,0) to (1,1) in the ROC space. 
    
\begin{figure}[h!]
    \centering
        \includegraphics[width=0.65\linewidth]{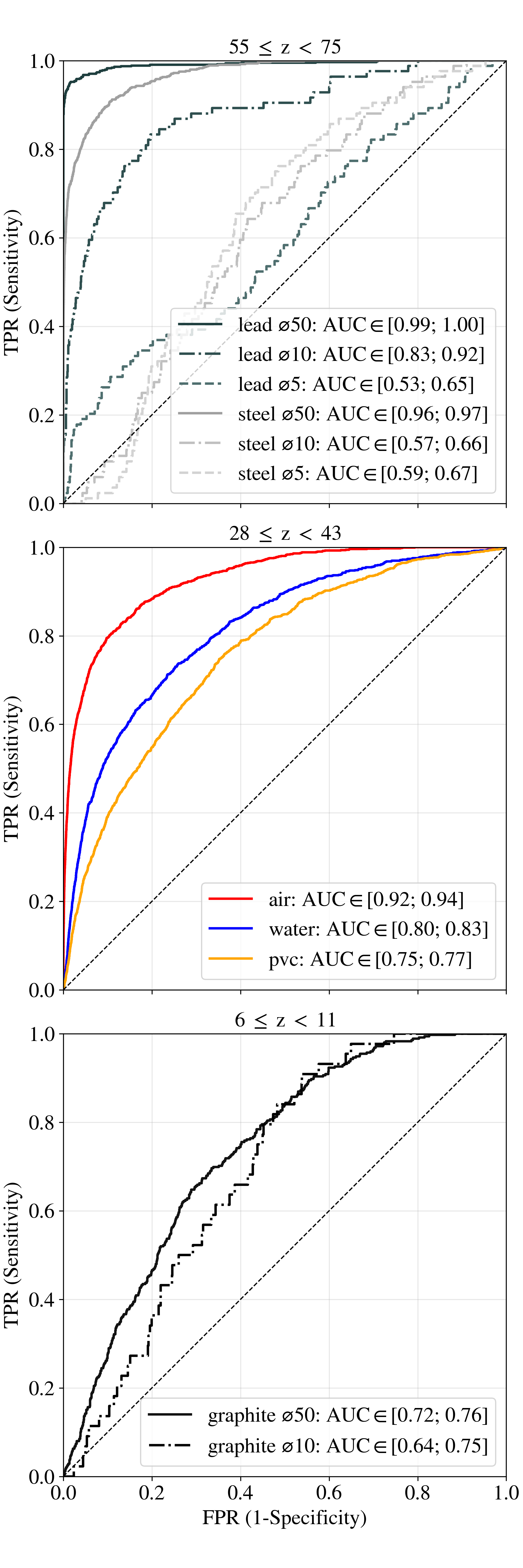}
        \caption{ROC curves along with their AUC obtained from the full simulation dataset for the different objects compared to concrete. The curves are grouped according to the corresponding object vertical coordinate range.}
        \label{fig:roc_curves}
    \end{figure}

    In other terms, for a targeted object, e.g., the lead bar, the AUC corresponds to the probability that a true lead voxel gets a higher score value than a concrete voxel. For objects with a radiation length larger than that of concrete — i.e., the cavities (air, water, PVC) — the scattering score is expected to be lower, rather than higher, than in the surrounding concrete. To preserve a consistent interpretation of the AUC across all object types, whereby values above 0.5 indicate effective discrimination, the sign of the score is reversed for cavities prior to the ROC computation; the AUC then correspondingly reflects the probability that a true cavity voxel gets a lower score value than a concrete voxel.
    In the following, discrimination performance is assessed in three steps: the overall AUC obtained from the full simulation dataset for each object (\textit{AUC estimation}); an evaluation of detection potential under short-exposure, "early" acquisition conditions (\textit{Early detection}); and the evolution of AUC with increasing exposure time, used to determine saturation levels and the corresponding experimental exposure required for optimal identification (\textit{AUC evolution}).
    \paragraph{AUC estimation}
    For this study, two simulated datasets were built: dataset (A), containing about $7.7\times10^6$ scattered events through the mock barrel schemed in Fig. \ref{fig:barrel_content_simu}, and dataset (B), of the same size, obtained on a full-concrete barrel. Only the barrel volume was selected, excluding the air in the VOI corners. For each object, the positive voxel set is defined as all voxels of dataset (A) located within the true geometrical volume of that object, as known from the simulation input. The corresponding negative voxel set consists of the voxels at the same spatial position and height range in dataset (B), where only concrete is present. Pairing positive and negative voxels by position and height range, in addition to the z-dependent score normalization described above, prevents any bias linked to the score's dependence on the crossed concrete thickness. The ROC curves obtained for all barrel objects compared to concrete are shown in Fig. \ref{fig:roc_curves} and grouped according to the three concerned z ranges, each annotated with the resulting AUC. The AUC 95\% confidence intervals (CI) were estimated by bootstrapping on the positive and negative voxel sets.
    
    The best detection performances are logically achieved for the largest \O50-mm lead bar with a maximal AUC value between [0.99; 1.00]. The \O50-mm steel bar is just below maximum between [0.96; 0.97]. 
    Among the two smaller diameter inclusions, only the \O10-mm lead bar reaches a significant detection level with final AUC $\in [0.83; 0.92]$. The other three metallic inclusions do not reach such a level and remain below $0.7$, which is backed by the observations in corresponding images (for z $\in[55;75]$) in Fig.\ref{fig:score_simu}. For the three cavities, only half of their voxels in height were selected to prevent potential score overlaps with metallic inclusions above. Their AUC estimations find themselves sorted according to the $X_0$ contrast with concrete: the air and water cavities get significant AUC intervals, [0.92; 0.94] and [0.80; 0.83] respectively, while the value for PVC is just above 0.75. At the bottom of the barrel, as expected from the minimal contrast with concrete, the graphite bars detection does not perform better, and AUC values remain at the same level.

    \paragraph{Early detection} To assess the potential of an "early" detection, a sub-sample of 10$^5$ events was considered, equivalent in real detection conditions to a six-hour bench exposure. The two simulation datasets (A) and (B) were divided into 77 realisations for AUC 95\% CI estimation. 
    The results of the early AUC estimation are presented in the fourth column of Tab.\ref{tab:SummaryAUC}. Only the \O50-mm lead bar was found to provide a significant AUC between [0.79; 0.81]. 
    
\begin{figure}[ht]
    \centering
        \includegraphics[width=0.75\linewidth]{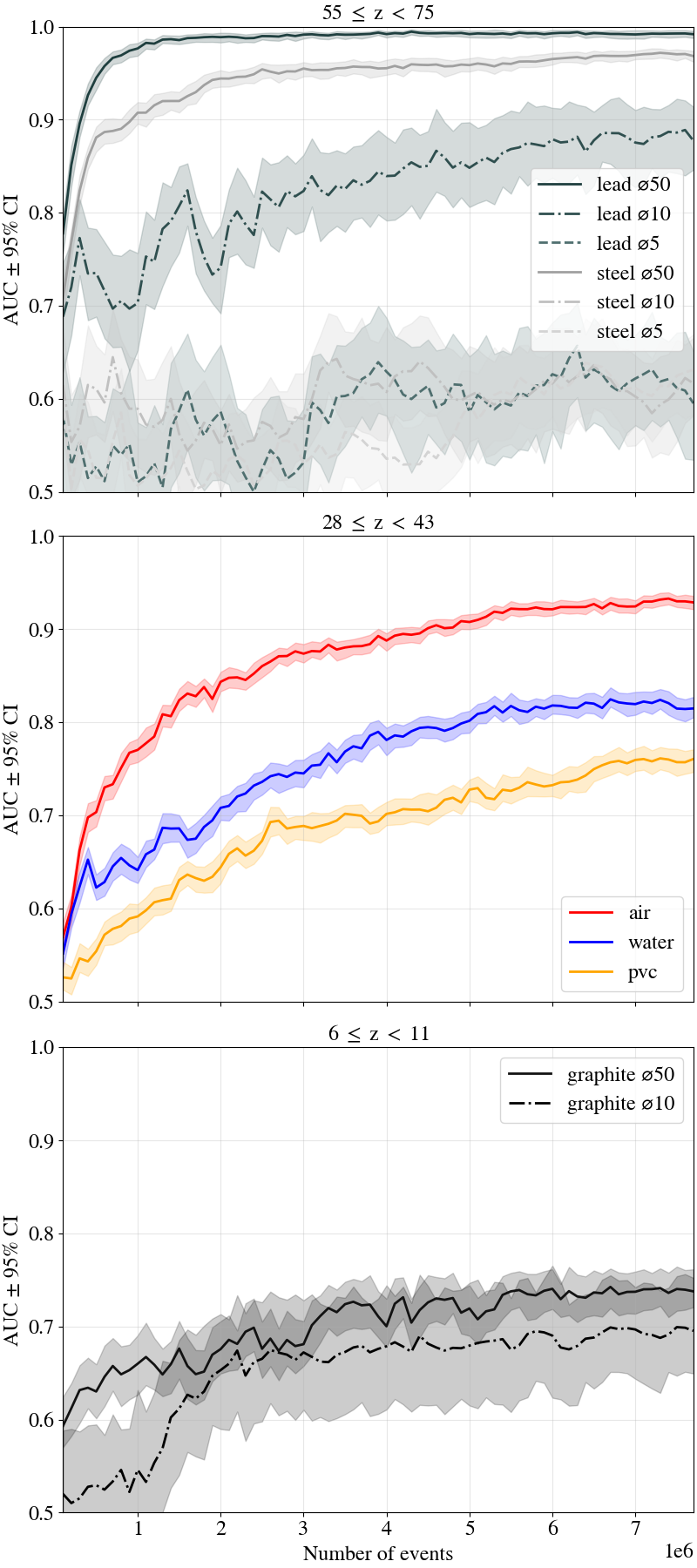}
        \caption{AUC as a function of number of events obtained from full simulation dataset for the different objects compared to concrete, drawn with their 95\% confidence interval estimated by bootstrapping. The curves are grouped according to the corresponding object height range.}    
        \label{fig:auc_curves}
    \end{figure}

    \paragraph{AUC evolution} Lastly, to better estimate the required exposure to reach optimal identification, the objects' AUC were estimated by increments of $10^5$ events. The resulting evolution curves are plotted in Fig.\ref{fig:auc_curves}. For the curves reaching a saturation level, an AUC asymptotic value was determined through a least-squared fit of data using function $f(N) = a+b(1+\exp(-N/c))$, with $N$ the number of events, $(a,b,c)$ as free parameters, and $ \lim\limits_{N\to\infty} f(N) = a+b = \text{AUC}_{\text{asymp}}$. The number of events to reach $95\%$ of AUC$_{\text{asymp}}$ can then be considered as the detection saturation level and gives an estimate of the experimental exposure time $\Delta t_{\text{sat}}^{\text{exp}}$ required for optimal identification. Results are summarized in Tab.\ref{tab:SummaryAUC}.
    Both metallic inclusions of \O50-mm reach a quasi-perfect identification level, in 2.1 days of exposure for lead and 4.8 days for steel. For the \O10-mm lead bar, the saturation level was extrapolated to 30.2 days. The other inclusions mean values remain globally low (AUC$\lesssim 0.65$) and do not achieve relevant saturation levels to be estimated in these conditions. Regarding the evolution of AUC for cavities, the air reaches a plateau below 0.94 after 11.0 days, while the PVC and water saturation levels are extrapolated at 22.5 and 22.8 days. For the two graphite objects, the AUC saturation remains below 0.75 for the \O50-mm bar, which coincides visually with the absence of clear object contrast in the three bottom images in Fig.\ref{fig:score_simu}.
    
    \input{Figures/tab1_roc_summary}

    \subsection{Anomaly detection in real data}
    \label{sec:AnomalyDetection} 
\begin{figure}[h!]
        \centering
        \includegraphics[width=0.9\linewidth]{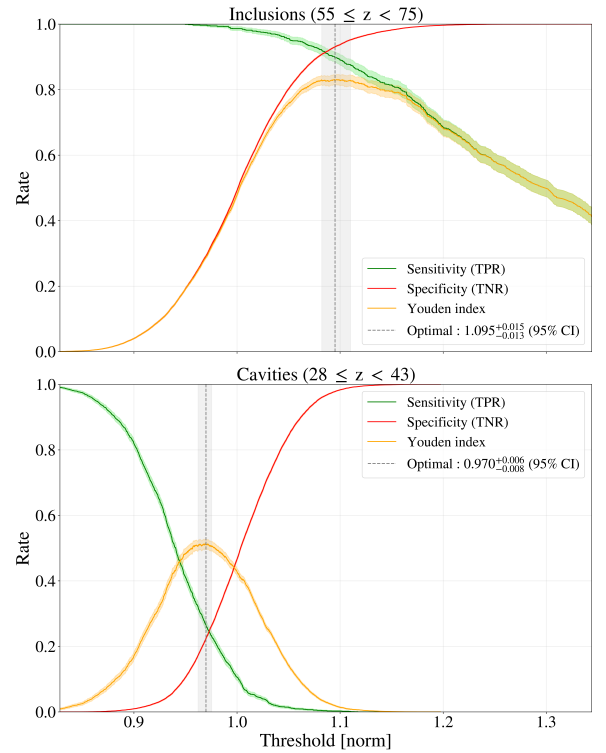}
        \caption{Sensitivity (green), specificity (red), and Youden index $J$ (orange) plotted as a function of threshold values, along with the optimal threshold at $\max(J)$ estimation in grey. The optimal threshold CI was estimated by bootstrapping on the positive and negative voxels. The plotted quantities are the ones that correspond to the optimal threshold in dash grey, with their binomial error (95\% CI).}    
        \label{fig:threshold}
    \end{figure}

    \paragraph{Threshold selection}
    To identify inclusions and cavities in the real data model, optimal score thresholds are first estimated from an ROC analysis conducted on simulation data. The maximum of Youden’s index ($J = \text{Sensitivity} + \text{Specificity} - 1$) is identified separately for inclusions and cavities over relevant height ranges (Fig.~\ref{fig:threshold}). These values define conservative global thresholds used to detect, respectively, local maxima for inclusions and local minima for cavities in the real model.
    \paragraph{2D extrema detection}
    Based on these thresholds, local two-dimensional extrema are detected on a series of horizontal projections computed along the vertical axis. To improve robustness to noise, projections are obtained using a sliding window along the Z-axis, applying a local minimum (for cavities) or maximum (for inclusions) operator over a fixed window width. On each projection, a minimum lateral separation of five voxels is enforced between detected extrema to avoid multiple detections of the same anomaly.
    \paragraph{3D tracking across depth}
    Local extrema detected on successive projections are then associated across depth using a nearest-neighbour tracking scheme based on a KD-tree \cite{virtanen_scipy_2020}. This tracking is performed sequentially along the Z-axis, enforcing vertical continuity and a bounded lateral displacement (less than three voxels) between consecutive slices. Each resulting track thus represents a candidate vertical structure. A three-dimensional candidate position is subsequently estimated by computing a score-weighted barycentre of the tracked extrema, giving higher weight to stronger local extrema.
    This directional tracking approach in Z is particularly well suited to the present physical case, where anomalies are expected to exhibit a predominantly vertical, cylindrical geometry with limited radial extent, while allowing for moderate lateral drift.
    \paragraph{Height and radius refinement}
    The height of each anomaly candidate is then refined by analysing the vertical score profile at the estimated lateral position. Consecutive voxels satisfying the threshold criterion are selected, allowing for a tolerance of one-voxel gaps to account for local noise fluctuations. The radial extent is subsequently estimated at each valid height by iteratively expanding a circular ring centred on the candidate position, with a stopping criterion based on the fraction of voxels satisfying the threshold. The final anomaly radius is estimated as the median of the radii obtained across the valid height range, providing a robust measure against local outliers. 
   
\begin{figure}[!htbp]
    \centering
    \begin{minipage}[b]{0.48\textwidth}
        \centering
        \includegraphics[width=\linewidth]{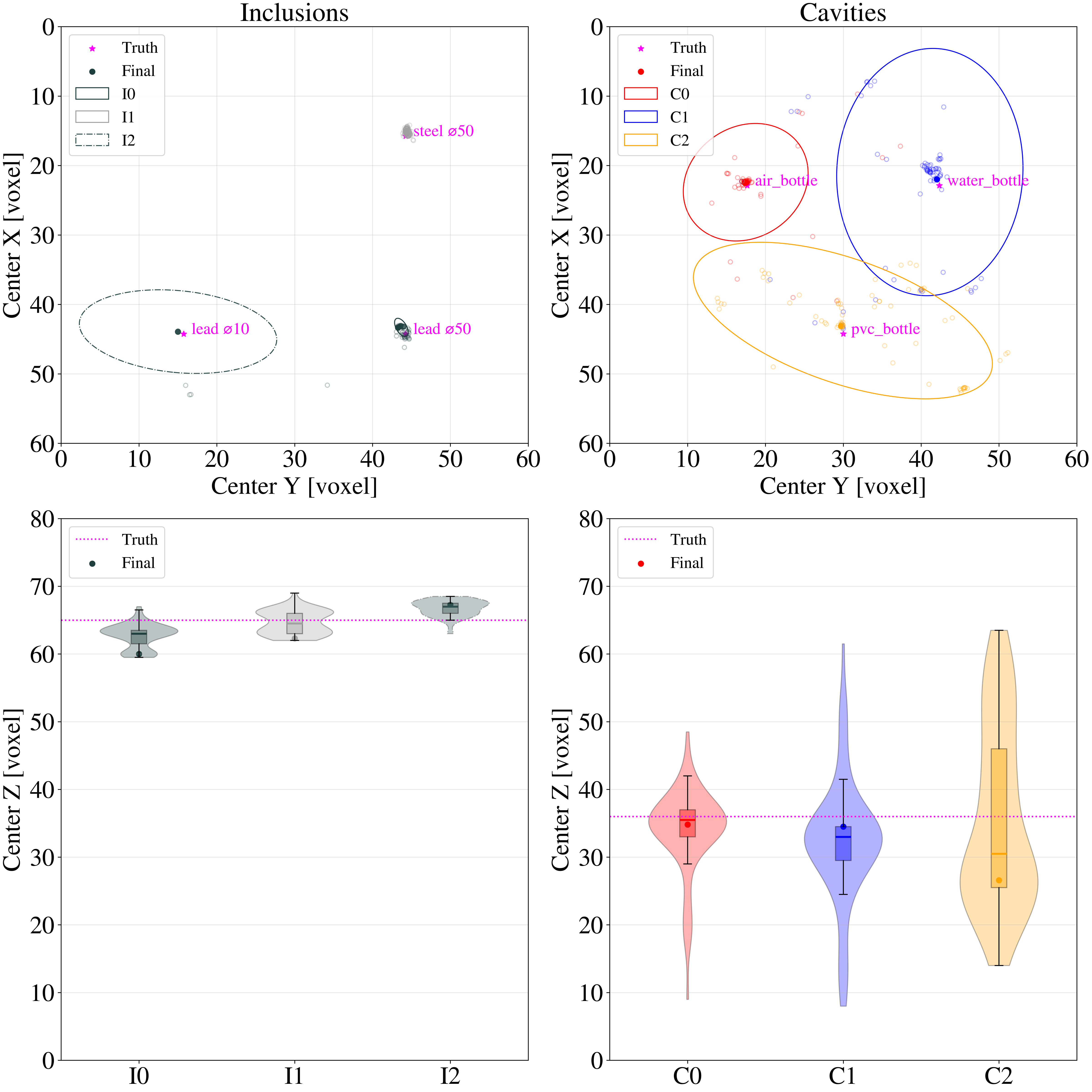}
        \caption{Estimates of the anomaly centre XY (ellipses) and centre Z (box plots) positions dispersion along the acquisition sequence for \textbf{simulation} data. The final estimates are highlighted with a solid circle marker.}
        \label{fig:ellipse_centre_xyz_simu}
    \end{minipage}
    \hfill
    \begin{minipage}[b]{0.48\textwidth}
        \centering
        \includegraphics[width=\linewidth]{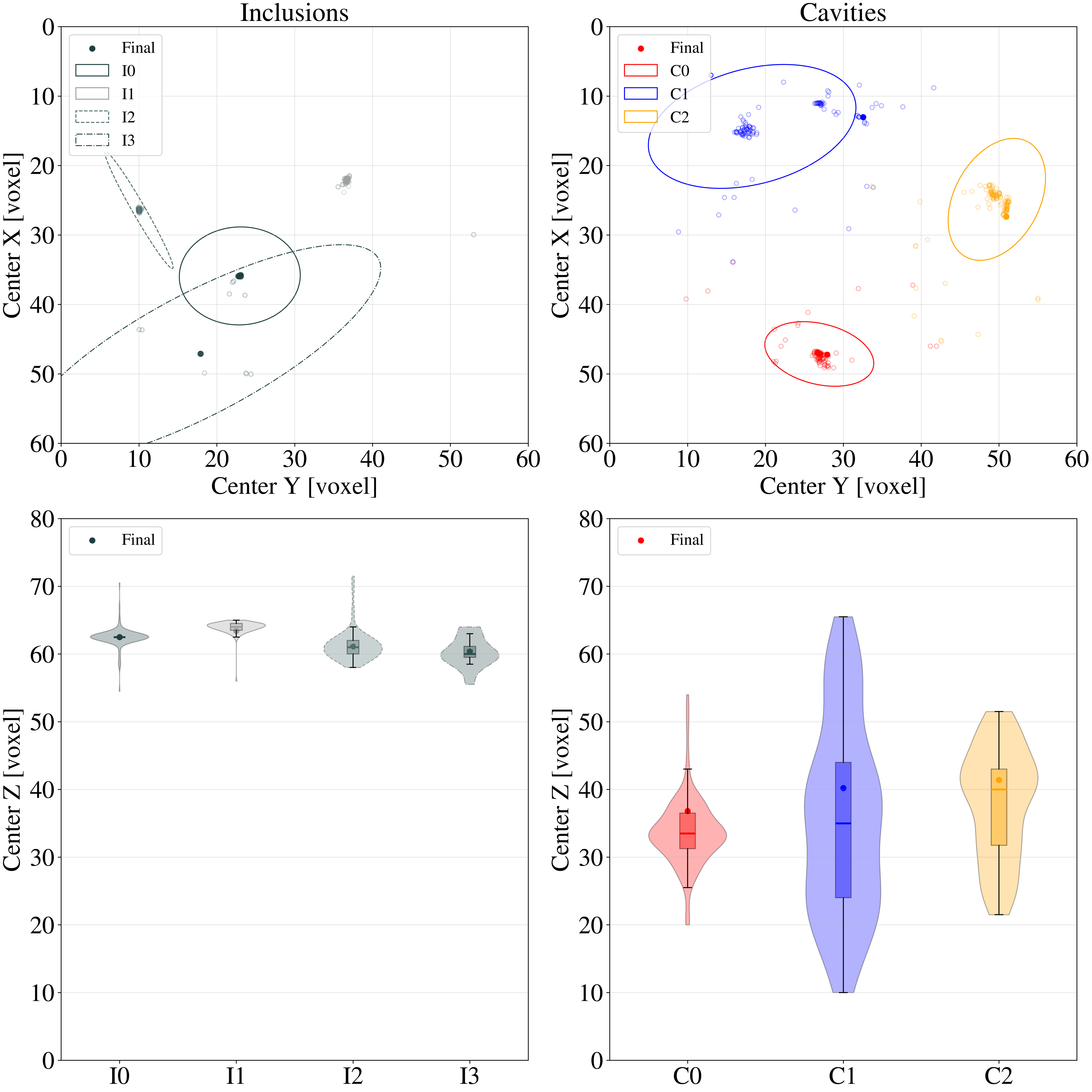}
        \caption{Estimates of the anomaly centre XY (ellipses) and centre Z (box plots) positions dispersion along the acquisition sequence for \textbf{real} data. The final estimates are highlighted with a solid circle marker.}
        \label{fig:ellipse_centre_xyz_real}
    \end{minipage}
    \label{fig:centre_xyz_anomaly}
    \end{figure}

\begin{figure}[!htbp]
    \centering
    \begin{minipage}[b]{0.48\textwidth}
        \centering
        \includegraphics[width=\linewidth]{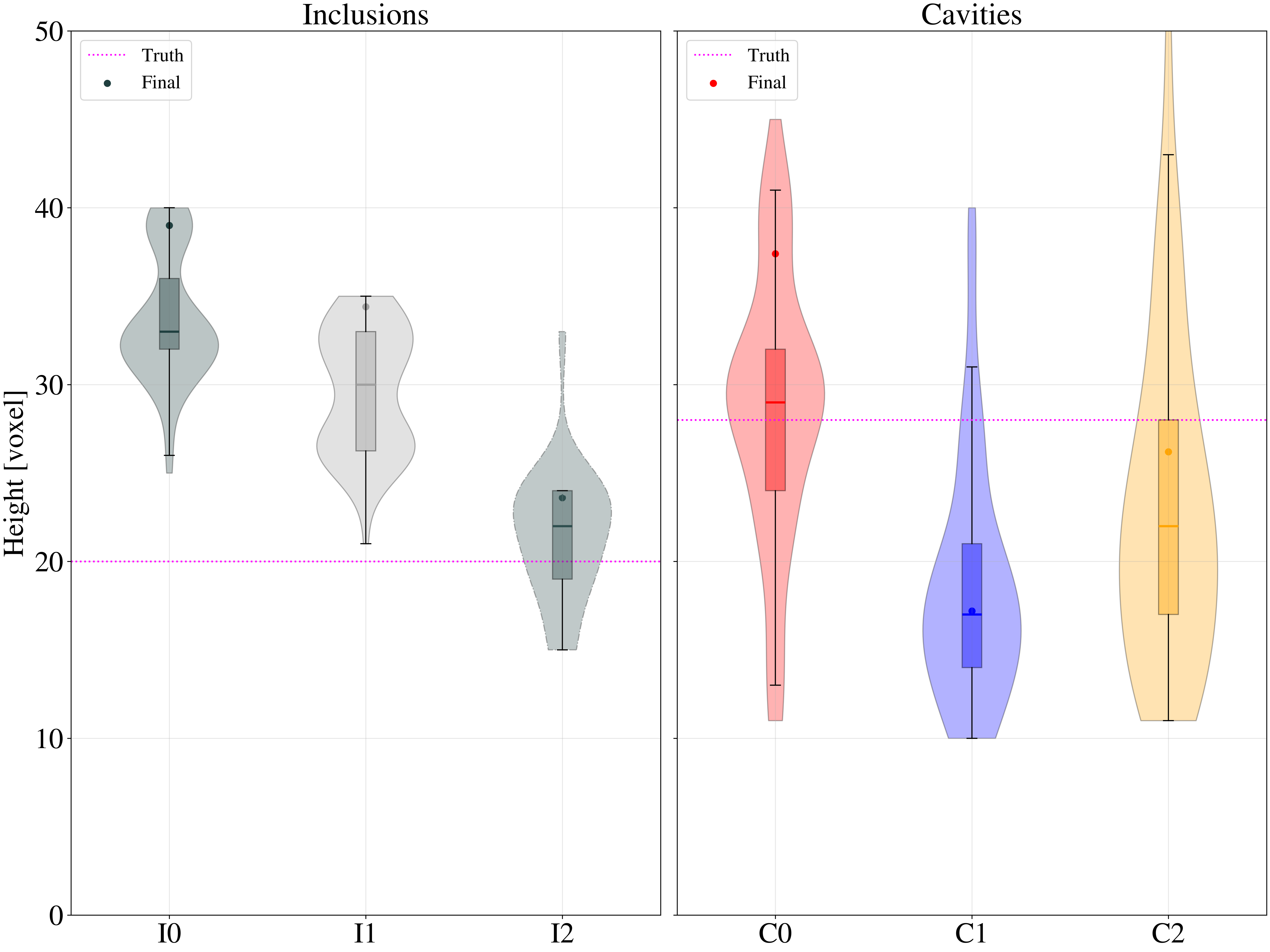}
        \caption{Height estimates of the different anomalies along the acquisition sequence for \textbf{simulation} data. The final estimates are highlighted with a solid circle marker.}
        \label{fig:height_anomaly_simu}
    \end{minipage}
    \hfill
    \begin{minipage}[b]{0.48\textwidth}
        \centering
        \includegraphics[width=\linewidth]{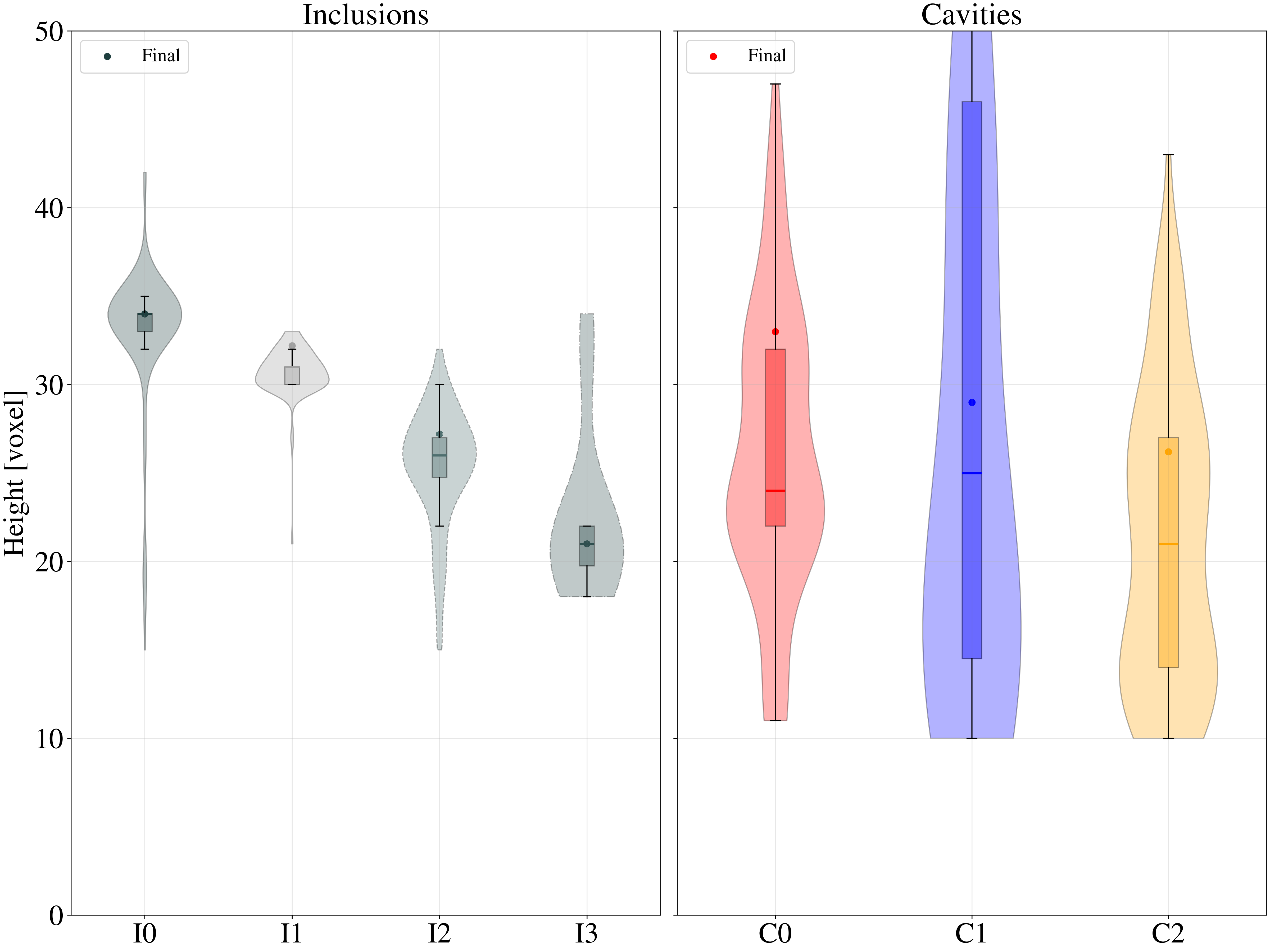}
        \caption{Height estimates of the different anomalies along the acquisition sequence for \textbf{real} data. The final estimates are highlighted with a solid circle marker.}
        \label{fig:height_anomaly_real}
    \end{minipage}
    \label{fig:height_anomaly}
    \end{figure}

   In the Fig.\ref{fig:ellipse_centre_xyz_simu} and \ref{fig:ellipse_centre_xyz_real}, we represent the empirical dispersion of the reconstructed anomaly centres over the full acquisition sequence. The drawn ellipses, centred on the median XY position with a $2\sigma$ width, summarize the variability of the anomaly reconstruction method across all event increments ($10^5$), affected by both statistical fluctuations and structural ambiguities in the model. The distribution of the centre Z is represented on a violin plot overlapped with a box plot featuring the median value and the inter-quartile range. 
    In simulation data, the final estimations of horizontal positions for all anomalies are overall exact with an absolute error inferior to 1.3 voxels for inclusions and 1.9 voxels for cavities (cf. Tab.\ref{tab:AbsErrorsAnomalyId}). The estimation of the vertical centre Z position is more imprecise, with maximal errors of 5.0 and 9.4 voxels, respectively, for the 50-mm lead inclusion and the PVC cavity. Furthermore, in Fig.\ref{fig:height_anomaly_simu} and \ref{fig:height_anomaly_real}, the final height estimates for inclusions are significantly overestimated for both simulation and real data. Relative to the nominal object extent along Z (20 voxels for the 50- and 10-mm lead bars and the 50-mm steel bar, and 28 voxels for the PVC, air, and water bottles), the height overestimation reaches approximately 95\% for the 50-mm lead bar, 72\% for the 50-mm steel bar, and 18\% for the 10-mm lead bar (cf. Tab.\ref{tab:AbsErrorsAnomalyId}). The three cavities are affected to markedly different extents: the air and water cavities show an overestimation of about 34\% and 39\%, respectively, while the PVC cavity is comparatively less impacted, at approximately 9\%.
    
    Moreover, a significant discrepancy is noticed in the shape reconstruction of the water cavity between the simulation and the real data. The cavity is indeed widely dispersed in the real model in Z and affected by two structural artefacts seen in the top centre of the XY position images (cf. Fig. \ref{fig:centre_xyz_anomaly}).   
    In the real dataset, a wider dispersion ellipse of the centre XY position of the I0 anomaly also differs from the simulation, matching the real \O50-mm lead bar region. The \O10-mm lead bar is known to be physically present in the region of reduced detection efficiency previously described (cf. Sec.\ref{sec:ResultsRealData}), yet it does not appear as a distinct feature in this region's images. An anomaly, labelled I3, is nonetheless localized there by the detection algorithm; however, given the elevated noise level in this region, this detection cannot be confidently attributed to a genuine identification of the 10-mm lead bar. The position ellipse of I3 overlaps with that of I0, whose own dispersion is broader than in simulation. We interpret this overlap, together with the broadening of the I0 ellipse, as evidence that both effects originate from the same underlying noise degradation affecting this region.

    \input{Figures/tab2_anomaly_detection}
 
\begin{figure}[!ht]
    \centering
    \includegraphics[width=\linewidth]{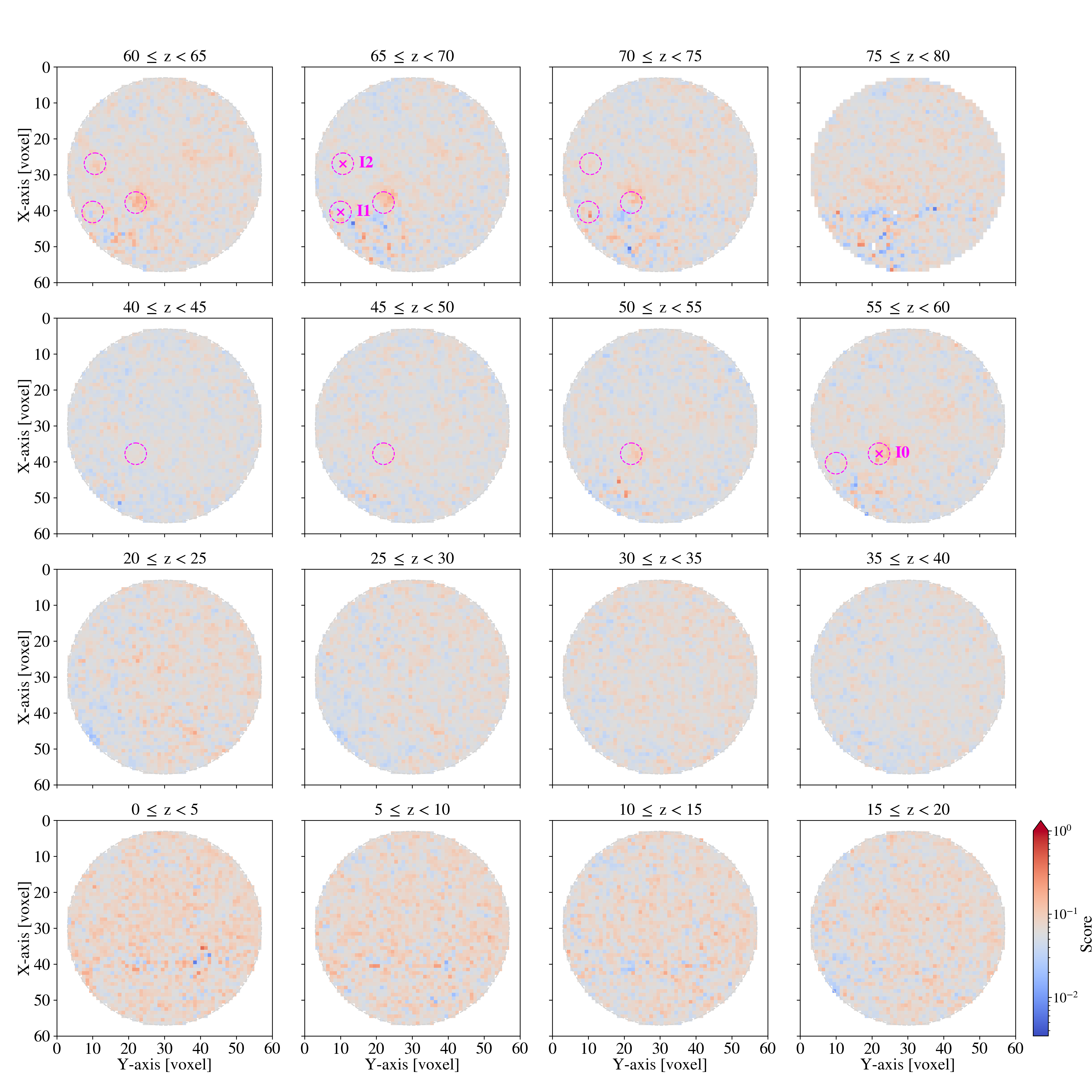}
    \caption{Early detection of inclusions in real data using the first $10^5$ events. The anomaly centres are marked with magenta crosses, along with estimated radii, and heights projected in the corresponding XY slices five voxels thick.}
    \label{fig:early_detection_real}
    \end{figure}

    Lastly, after analysing through the ROC study and AUC estimates in Sec. \ref{sec:RocStudy} the potential for an early detection at the scale of $10^5$ events, we explicitly show in Fig.\ref{fig:early_detection_real} the ability to perform the anomaly shape identification (inclusions only) in the real model projections using only the first $10^5$ events and an early threshold value estimated at $1.248 ^ {+0.027}_{-0.038}$ in the same way that presented before (cf. Fig. \ref{fig:threshold}). The \O50-mm lead bar, labelled I0 in the figure, is the first to create sufficient score contrast in the model along a height longer than twenty voxels and a three voxel radius ranging from $z\in[40;75]$, in accordance with the simulation AUC estimates in Tab.\ref{tab:SummaryAUC}. In addition to this main anomaly, another (I1) was reconstructed in the noise-dominated region of the projections in $z\in[50;75]$ not matching any of the introduced physical object, and a third one (I2) was found around the \O5-mm lead bar, which is a major difference with the simulation predictions where this anomaly showed very low AUC values both in the short and long run operation (cf. Tab.\ref{tab:SummaryAUC} and Fig.\ref{fig:auc_curves}). Furthermore, we noted that the \O50-mm steel inclusion shape starts to be reconstructed over $4.10^5$ and $6.10^5$ events in the simulation and the real model, respectively.

    \subsection{Discussion}
    \label{secDiscussion}
        
    In the previous section, angular statistics reconstruction (ASR) results obtained from both simulation and real datasets were used to evaluate the capability of MST to detect and characterize multiple materials of varying shapes, densities, and radiation lengths embedded in a concrete barrel structure, at a voxel resolution of 10 mm. Detection performances were investigated under both short- and long-exposure conditions.
    
    In simulation, receiver operating characteristic (ROC) analyses were first employed to assess material separability and to estimate optimal score thresholds for inclusions and cavities. These thresholds were subsequently used in a dedicated spatial reconstruction pipeline applied to both simulated and real data, allowing anomaly positions and shapes to be estimated from local extrema of the ASR score.
    Anomaly reconstruction relies on identifying local extrema on each horizontal slice of the model, followed by their association across successive depths using a nearest-neighbor tracking scheme based on a KD-Tree \cite{virtanen_scipy_2020}. This approach, characterized by a computational complexity of $\mathcal{O}(\log N)$, was selected for its directional nature along the Z-axis, which is well suited to vertically extended, cylindrical anomalies with limited lateral displacement. In contrast, isotropic clustering approaches such as DBSCAN \cite{Ester1996} do not explicitly enforce vertical continuity and were therefore deemed less appropriate for the present geometry.
    
    Among the eleven objects embedded in the concrete, six exhibited strong detection potential in the simulation-based ROC analysis, as reflected by their AUC values (\O50- and \O10-mm lead bars, \O50-mm steel bars, and air, water, and PVC cavities). These findings were further corroborated by the spatial reconstruction results. Under short exposure conditions ($10^5$ events, corresponding to approximately six hours of acquisition with the current detection efficiency), the \O50-mm lead inclusion was the first to be clearly identified, followed by the \O50-mm steel inclusion, which required approximately twice as many events to achieve comparable contrast.
    
    A notable discrepancy between simulation and real data concerns the detection of the \O5-mm lead bar in the experimental model. Its identification suggests that scattering-based reconstruction may be sensitive to sub-centimetre-scale inclusions. Several factors may contribute to this result, including a slight vertical inclination of the bar during assembly, which would effectively increase its projected radial cross-section, and/or a locally favorable signal-to-noise configuration within the detection setup.
    In particular, incorporating data-driven efficiency maps derived from the measured spatial response of each MultiGen detector into the GEANT4 simulation would allow a more faithful reproduction of the local artefacts observed in real data (Fig. \ref{fig:score_real}), and is expected to further improve the quantitative agreement between simulated and experimental scattering-score models.

    Overall, the ASR modelling demonstrates promising capabilities for anomaly identification in both simulated and real configurations. However, when applied to a vertically distributed set of cylindrical objects, a pronounced spreading of the scattering score along the Z-axis was observed, especially for high-$X_0$ objects (cavities). This effect impacts the estimation of both anomaly height and vertical position, potentially leading to structural ambiguities when multiple objects are located in close vertical proximity. In contrast, anomaly localization in the XY plane—particularly for dense inclusions—appears more stable, with limited lateral score spreading beyond the object radius, as reflected by the low dispersion of reconstructed XY centers. 
    
    The vertical spreading discussed above translates directly into the object height overestimation quantified in Sec. \ref{sec:AnomalyDetection} (cf. Tab.\ref{tab:AbsErrorsAnomalyId}), and consequently limits the achievable resolution on volume estimation for vertically compact or closely spaced inclusions. Notably, the relative impact of this blurring differs markedly between dense metallic inclusions and cavities and does not scale simply with object size, suggesting that both the material's radiation-length contrast and its vertical extent govern the resulting bias. A full translation of this effect into a Minimum Detectable Mass or Volume metric would require a dedicated analysis linking the AUC saturation levels (Tab.\ref{tab:SummaryAUC}) to these height-estimation statistics across a broader range of object sizes and placements, which we identify as an important direction for future work. 
    
    Beyond these geometrical ambiguities, the present results also raise an intrinsic boundary condition of MST. As quantified by the AUC estimates in Sec.\ref{sec:RocStudy}, the discrimination of graphite and, to a lesser extent, PVC from the surrounding concrete remains limited even under extended exposure, reflecting the comparatively weak density and atomic-number contrast probed by multiple Coulomb scattering alone for such materials. This constitutes a relevant limitation for applications targeting organic waste matrices or neutron-moderating compounds, for which a single scattering-based observable is expected to offer reduced discrimination power. Combining the scattering score with complementary muon observables, most notably absorption-based measurements, has recently been investigated as a means of improving the separation of high-$X_0$ (low-$Z$) materials from denser backgrounds (\cite{Vanini2018}, \cite{Chen2023}, \cite{Georgadze2024}, \cite{He2024}). 
    The present work focuses on a fully volumetric, scattering-based reconstruction. Extending the current pipeline to incorporate such combined observables is a promising direction to improve the bench's discrimination capability for lower-contrast, high-$X_0$ compounds, without departing from the three-dimensional reconstruction approach at the core of the present methodology.
    
    While the present MultiGen detectors are optimised for spatial resolution rather than timing, the intrinsic timing capabilities of the Micromegas technology could, in principle, be further exploited to access muon momentum information via Time-of-Flight (ToF) measurements. This possibility is illustrated by PICOSEC-Micromegas, a dedicated fast-timing architecture combining a Cherenkov radiator, a photocathode, and a bulk Micromegas amplification stage, which has demonstrated a time resolution below 25 ps for minimum-ionising particles in single-pad prototypes \cite{Kordas2020}, with this performance shown to be preserved when scaling to larger-area, multi-channel devices \cite{Kallitsopoulou2024}. It should be stressed that PICOSEC constitutes a distinct architectural evolution within the Micromegas family, developed independently from the multiplexed resistive design of the MultiGen detectors used in the present bench, and does not represent a capability of the current system. Access to individual muon momentum, presently treated as a fixed parameter in Eq. \ref{eqTheta0}, would help disentangle the respective contributions of momentum and crossed material to the observed scattering angle, and constitutes a motivation for exploring dedicated fast-timing Micromegas developments in future iterations of the present bench.

    More broadly, for application to genuine waste packages, hybrid approaches coupling MST with neutron-based interrogation techniques could provide additional sensitivity to organic and hydrogenous materials: passive neutron and gamma-ray detection has already been combined experimentally with MST for nuclear material detection \cite{Kamaev2019}, while prompt and delayed gamma neutron activation analysis has been proposed to access the elemental composition of waste drums, including light and non-radioactive constituents that remain largely inaccessible to muon-only techniques \cite{Coquard2023}.
    
    Future work will therefore focus on (i) further characterizing the vertical ambiguities identified above, including comparison with alternative statistical reconstruction methods offering improved handling of vertical localization and uncertainty quantification, (ii) translating these results into quantitative Minimum Detectable Mass/Volume metrics, and (iii) exploring, for application to real waste packages, the integration of complementary muon-based (scattering/absorption, ToF) and neutron-based observables to extend material discrimination toward high-$X_0$, low-contrast, and neutron-moderating compounds.


    
    \section{Summary and Conclusion}

    This study demonstrates the feasibility and performance of cosmic-ray MST for the non-destructive characterization of complex, shielded structures, using a mock nuclear waste barrel as a representative case. The deployment of a dedicated 1 m$^{2}$ experimental test bench equipped with resistive multiplexed Micromegas detectors enabled stable and high-precision muon tracking under conditions relevant to nuclear waste inspection.

    The experimental campaign successfully imaged a 205-L metallic barrel filled with concrete and containing a variety of low- and high-$X_0$ materials, including lead, steel, graphite bars, as well as PVC, water, air cavities. Using the Angle Statistics Reconstruction (ASR) algorithm, a consistent agreement between Monte Carlo simulations and experimental data was observed, validating the modelling of material-dependent scattering signatures and the reconstruction approach.
    
    Monte Carlo simulations were first employed to quantitatively assess material discrimination performance using ROC analysis and AUC metrics, and to define objective identification thresholds. These thresholds were subsequently applied to experimental data, enabling not only the identification but also the spatial localization and geometric reconstruction of internal anomalies within the barrel at a voxel resolution of 10-mm. Dense metallic inclusions such as lead and steel exhibited high discrimination performance within acquisition times of a few days, while cavities also showed strong and stable contrast at longer exposure times.
    

    Beyond detection performance, this highlights two main current limitations of the scattering-only approach. First, a vertical spreading of the scattering score affects the reconstruction of anomaly height and depth in vertically extended configurations, limiting the achievable resolution on volume estimation for compact or closely spaced inclusions. Second, the discrimination of high-$X_0$ materials such as graphite or PVC from the concrete medium remains marginal even under extended exposure. 
    Addressing these limitations is essential for improving the robustness of anomaly separation and material discrimination in densely packed or more complex waste containers. 

    Future developments will focus on mitigating the vertical spreading through improved reconstruction strategies and uncertainty handling, translating these results into quantitative Minimum Detectable Mass/Volume metrics, and extending the present methodology toward the characterization of high-$X_0$, low-contrast, and neutron-moderating compounds in larger scale systems and more complex configurations.

    By establishing a continuous workflow linking simulation-based performance quantification to practical anomaly localization in real data, this study provides a validated framework for the application of Micromegas-based MST to non-invasive inspection of nuclear waste packages.
    
    \phantomsection
    \section*{Acknowledgments} 
    
    
    The authors would like to express their gratitude to the Orano company for funding this project in the UDD$@$Orano (\textit{Usine De Demain at Orano}) framework, supported by the French government. Their support has been instrumental in the development and execution of this research. We also acknowledge the contributions of our colleagues and collaborators who provided valuable insights and assistance throughout the study.
    
    \phantomsection
    
    \section*{Declaration of Generative AI Use} 
    During the preparation of this work, the authors used Grammarly for the purpose of language editing and grammar checking only. The tool was used to improve the readability and clarity of the text. After using this service/tool, the authors reviewed and edited the content as needed and take full responsibility for the content of the published article. No AI tool was used to generate scientific data, perform analysis, or formulate conclusions.
    
    
    \phantomsection
    \bibliography{biblio}
    
    
    \end{document}

%% file: Figures/tab1_roc_summary.tex
\begin{table*}[ht!]
\centering
\small
\begin{tabular}{w{c}{1.2cm} w{c}{1.2cm} w{c}{1.cm} w{c}{1.cm} w{c}{2.cm} w{c}{1.5cm} w{c}{1.3cm} w{c}{1.3cm}} 
\hline
Z [mm]&Material& $X_0$ [cm] & \O [mm] & AUC$_{\text{early}}$ & AUC$_{\text{asymp}}$ & N$_{\text{sat}}$ & $\Delta t_{\text{sat}}^{\text{exp}}$ [\text{days}] \\
\hline
\hline
\multirow{3}{*}{[550;750]}
& Lead & 0.56  & 50 & [0.79; 0.81] & 0.99 (0.01) & 8.17e+05 & 2.1\\  
\cline{4-8}
&  & &  10 & [0.60; 0.63] & 0.92 (0.01) & 1.15e+07 & 30.2\\
\cline{4-8}
&  & & 5  & [0.58; 0.61] & - & - & -  \\ 
\cline{2-8}

\multirow{3}{*}{} 
& Steel & 1.76 &  50 & [0.68; 0.69] & 0.96 (0.01) & 1.85e+06 & 4.8 \\
\cline{4-8}
&  & & 10 & [0.57; 0.59] & - & - & - \\
\cline{4-8}
& & & 5  & [0.56; 0.59] & - & - & - \\ 
\hline

\multirow{3}{*}{[280;430]}
& Air & 3.1 $10^4$ & 85 & [0.58; 0.59] & 0.94 (0.01) & 4.19e+06 & 11.0\\
\cline{2-8}
& Water & 36.1  & 85 & [0.54; 0.55] & 0.85 (0.01) & 8.59e+06 & 22.5\\
\cline{2-8}
& PVC & 19.6 & 85 & [0.55; 0.56] & 0.77 (0.01) & 8.69e+06 & 22.8\\
\hline 

\multirow{2}{*}{[60;110]}
& Graphite & 19.4 & 50 & [0.56; 0.57] & 0.75 (0.01) & 9.34e+06 & 24.5\\
\cline{4-8}
& & & 10 & [0.54; 0.55]  & 0.70 (0.01) & 4.66e+06 & 12.2\\
\hline

\end{tabular}
\caption{Summary of ROC performances presenting early AUC value for $10^5$ events with 95\% CI, the asymptotic AUC value, along with N$_{\text{sat}}$ the number of events to reach 95\% of AUC$_{\text{asymp}}$ and corresponding experimental exposure $\Delta t_{\text{sat}}^{\text{exp}}$.} 
\label{tab:SummaryAUC}
\end{table*}

%% file: Figures/tab2_anomaly_detection.tex
\begin{table}[h!]
\centering
\small
\begin{tabular}{w{c}{1.3cm} w{c}{1.2cm} w{c}{.5cm} w{c}{1.cm} w{c}{1.4cm} w{c}{1.cm}w{c}{1.2cm}}
\hline
Anomaly & Material & \O & Label & Center XY & Center Z & Height\\
\hline
\hline
\multirow{3}{*}{Inclusions} 
& Lead &  50 & I0 & 1.3 & 5.0 & 19.0 \\  
\cline{3-7}
& &  10 & I2 & 0.8 & 2.3 & 3.6 \\  
\cline{2-7}
& Steel & 50 & I1 & 0.6 & 2.7 & 14.4 \\  
\hline
\multirow{3}{*}{Cavities} 
 & Air & 85 &  C0  & 0.6 & 2.7 & 9.4 \\  
 \cline{2-7}
 & Water & 85 &  C1  & 1.4  & 2.7 & 10.8\\
 \cline{2-7}
 & PVC & 85 &  C2 & 1.9  & 9.4 & 2.6 \\
\hline
\end{tabular}
\caption{Final absolute errors on anomaly centre position and height in simulation data, in voxel units.} 
\label{tab:AbsErrorsAnomalyId}
\end{table}